\documentclass[12pt]{iopart}

\usepackage{iopams}

\usepackage{graphicx}

\begin{document}

\title[]{Group-III assisted catalyst-free growth of InGaAs nanowires and the formation of quantum dots}

\author{Martin Hei\ss$^{1,2}$, Bernt Ketterer$^{1}$, Emanuele Uccelli$^{1,2}$, Joan Ramon Morante$^{3,4}$, Jordi Arbiol$^{5}$, Anna Fontcuberta i Morral$^{1,2}$} 
\address{$^1$ Laboratoire des
Mat\'{e}riaux Semiconducteurs, Institut des Mat\'{e}riaux, Ecole
Polytechnique F\'{e}d\'{e}rale de Lausanne, Switzerland}

\address{$^2$ Walter Schottky Institut and
Physik Department, Technische Universit\"{a}t M\"{u}nchen, Am
Coulombwall~3, D-85748 Garching, Germany}

\address{$^3$ Departament
d'Electr\`{o}nica,Universitat de Barcelona, E-08028 Barcelona, CAT,
Spain}

\address{$^4$ Catalonia Institute for Energy Research (IREC),
E-08019 Barcelona, CAT, Spain}
\address{$^5$ Instituci\'{o} Catalana de Recerca
i Estudis Avan\c{c}ats (ICREA) and Institut de Ci\`{e}ncia de
Materials de Barcelona, CSIC, 08193 Bellaterra, CAT, Spain}

\ead{anna.fontcuberta-morral@epfl.ch}

\begin{abstract}
Growth of GaAs and $\mathrm{In}_\mathrm{x}\mathrm{Ga}_\mathrm{1-x}\mathrm{As}$ nanowires by the group-III assisted Molecular Beam Epitaxy growth method is studied in dependence of growth temperature, with the objective of maximizing the indium incorporation. Nanowire growth was achieved for growth temperatures as low as 550\,$^\circ\mathrm{C}$. The incorporation of indium was studied by low temperature micro-photoluminescence spectroscopy, Raman spectroscopy and electron energy loss spectroscopy. The results show that the incorporation of indium lowering the growth temperature does not have an effect in increasing the indium concentration in the bulk of the nanowire, which is limited to 3-5\,\%. For growth temperatures below 575\,$^\circ\mathrm{C}$, indium rich regions form at the surface of the nanowires as a consequence of the radial growth. This results in the formation of quantum dots, which exhibit extremely sharp luminescence. 
\end{abstract}

\section{Introduction}

Nanowires are filamentary crystals with diameters in the order of few nanometers. Thanks to their peculiar shape and dimensions, they are a great promise for many technological advances in this century in diverse areas such as biosensing, energy harvesting and optoelectronics \cite{Cui2000,Cui2001,Samuelson2004,Wang2005,Kayes2005}. For the full deployement of such technology, control over the nanowire structure and composition at the nanometer scale is essential. In this frame, nanowire based heterostructures have been broadly studied \cite{Cui2001,Samuelson2004,Cornet2007,Dick2007}. 
Among the various compound semiconductors, InGaAs is considered to be one of the suitable materials as a transistor channel due to its low electron effective mass \cite{Vurgaftman2001}. Additionally, InGaAs/GaAs based quantum wells and dots constitute the ideal material system for infrared detectors and single photon emitters. Moreover, the nanowire geometry has turned to be ideal for all these applications as it enhances the functionality by allowing a better coupling with the electromagnetic radiation\cite{Borgstrom2005}.

Nanowires are commonly obtained through the vapor-liquid-solid, VLS, (or vapor-solid-solid, VSS) method, where gold is used as a catalyst that preferentially gathers and decomposes the precursors \cite{Dick2005,Paladugu2008}. As a result of the concerns that gold raises in the area of semiconductor technology, group III assisted growth has received an increased attention \cite{Yazawa1991,Novotny2005,Fontcuberta2008,Plissard2010}. This method is also compatible with the fabrication of heterostructures for example by combining the pairs InGaAs/GaAs \cite{Uccelli2010,Heiss2009} or zinc-blende/wurtzite crystal phases \cite{Zardo2009,Spirkoska2009}. 

InGaAs and InAs quantum dots have been obtained in GaAs nanowires grown by the VLS method (Au-assisted) \cite{Gunawan2010}. Investigations on the optical properties and interface sharpness indicate that the InAs/GaAs system is more favorable than the InGaAs/GaAs \cite{Paladugu2008,Panev2003,Krogstrup2009}. The synthesis of InGaAs nanowires with the Au-assisted method has been demonstrated by various groups. Martelli \textit{et al.} varied the growth temperature down to 480\,$^\circ\mathrm{C}$, and reached an indium content up to 22\,\% \cite{Jabeen2008,Jabeen2009}. However, the concentration turned not to be homogeneous along the nanowire axis. Such inhomogeneities have been attributed to the longer diffusion length of indium with respect to GaAs and have equally been observed by other groups \cite{Cornet2007,Jasik2009}. InGaAs nanowires have also been obtained by selective area epitaxy. There, the difference in diffusion length between gallium and indium results in a different indium incorporation depending on the distance between the nanowires \cite{Sato2008}. Up to date, it is still controversial to what extent homogeneous InGaAs nanowires can be obtained and what is the maximum indium concentration \cite{Sudfeld2006}. 
The synthesis of InGaAs nanowire and InGaAs/GaAs nanowire heterostructures by the group-III assisted method has been demonstrated recently \cite{Heiss2009}. In this case, the nanowire growth was performed under growth conditions optimized for the growth of GaAs nanowires. It was argued that due to the high growth temperatures, the incorporation of indium was limited to a few percent. It is expected that a higher incorporation of indium may be reached by lowering the growth temperature. At the same time, previous studies have shown that the sticking coefficient of gallium on SiO$_2$ increases as the substrate temperature is decreased and becomes close to unity for temperatures lower than 565\,$^\circ\mathrm{C}$ \cite{Heiss2008}. Unfortunately, the optimum growth temperature range for the growth of $\mathrm{In}_\mathrm{x}\mathrm{Ga}_\mathrm{1-x}\mathrm{As}$ compounds corresponds to temperatures below 520\,$^\circ\mathrm{C}$ \cite{Shen1990}. Indeed, by lowering the growth temperature down to the range of 400-505\,$^\circ\mathrm{C}$, pure InAs growth was obtained \cite{Kobelmueller2010,Colombo2007}. The challenge in catalyst-free growth is then, to find conditions where gallium still participates in the catalyst-free growth and where the incorporation of indium is enhanced. The first condition would normally require a temperature growth between 600 and 630\,$^\circ\mathrm{C}$, while the second the lowest growth temperature possible. In this work we study the influence of the growth temperature on the nanowire growth and the implications in terms of indium incorporation and optical properties.

\section{Experimental details} 
\label{ingaas:growthtempdep}
\begin{figure*}[tb!]
	\centering
		\includegraphics[width=\textwidth]{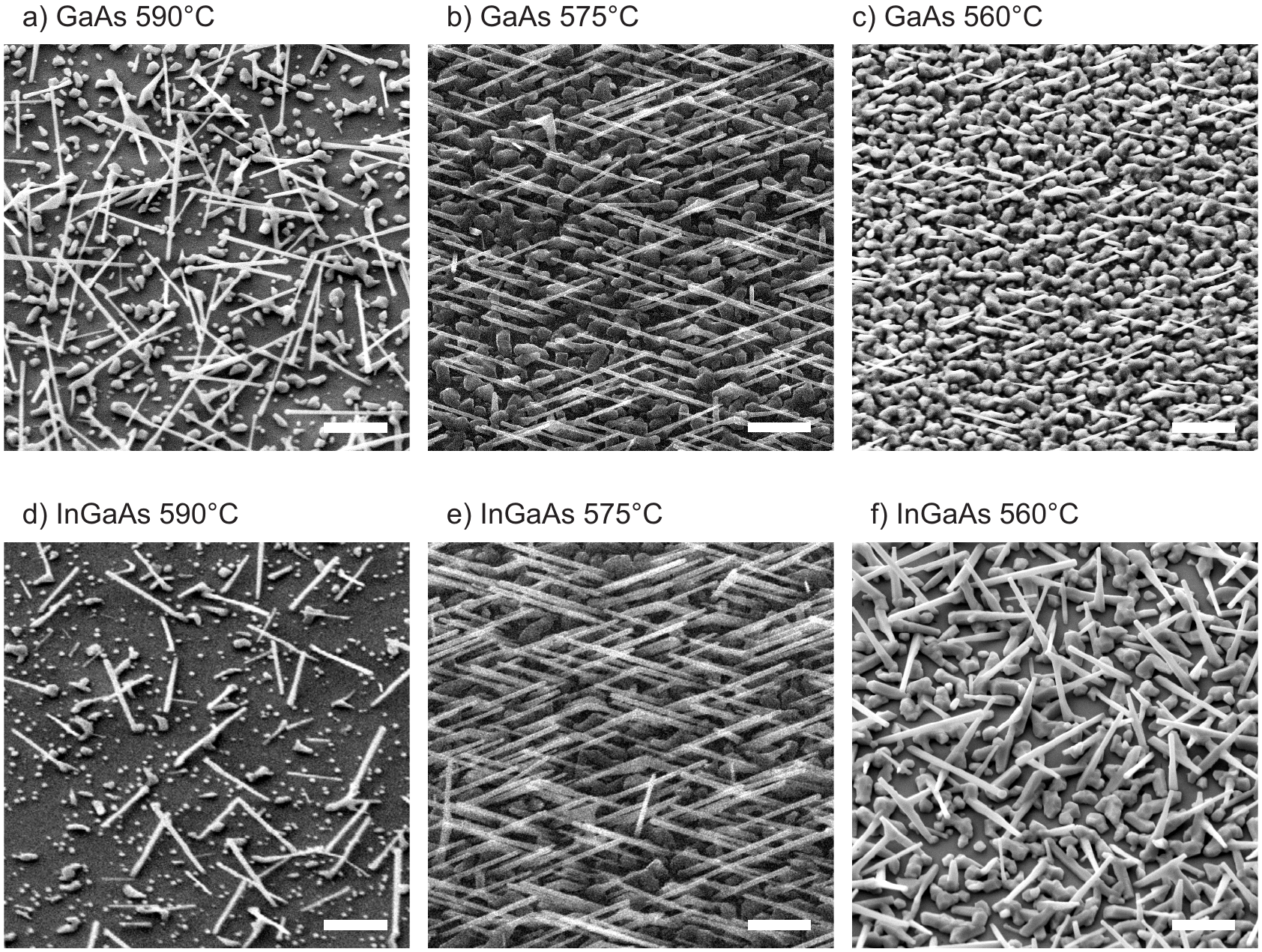}
	
	\caption{\label{ingaas:growthtempseries}Representative tilted view (30$^{\circ}$) SEM micrographs of GaAs and $\mathrm{In}_\mathrm{x}\mathrm{Ga}_\mathrm{1-x}\mathrm{As}$ nanowire samples grown at substrate temperatures of 560-590\,$^\circ\mathrm{C}$ at otherwise identical conditions for the same growth times. The scale bar in each of the micrographs corresponds to 1\,$\mu$m.}
	
\end{figure*}

The nanowires were grown by the group-III assisted method by molecular beam epitaxy as detailed elsewhere \cite{Fontcuberta2008,Heiss2009,Colombo2008}. For this, (001) oriented GaAs substrates covered with a thin layer of SiO$_2$ were used. A systematic growth temperature series where gallium or both gallium and indium was supplied during the entire growth process was realized. The growth temperature series for pure GaAs nanowires grown under the same conditions is used as a reference. The substrate temperature was varied in the range of 550-590\,$^\circ\mathrm{C}$. The indium and gallium rates were fixed at 0.045\,{\AA\,s$^{-1}$} and 0.2\,{\AA\,s$^{-1}$}, respectively. The As$_4$ beam flux was set to $8.8\cdot10^{-7}\,\mathrm{mbar}$ at a constant growth time of 5400\,s for all samples. 
Further details on the growth procedure can be found in \cite{Colombo2008}.
The morphology of the nanowires was studied by scanning electron microscopy (SEM). The composition and crystal structure of the nanowires was studied by high resolution electron microscopy (HRTEM), electron energy loss spectroscopy (EELS) and Raman spectroscopy using the 488\,nm line of an Ar$^{+}$Kr$^{+}$-laser focused to a diffraction limited spot (NA=0.75, power density 50\,kW\,cm$^{-2}$). The optical properties of single nanowires were studied by photoluminescence spectroscopy at 4.2\,K with a confocal microscope. The photoluminescence was excited with a 780\,nm diode-laser focused to a diffraction limited spot (NA=0.65). For this, the nanowires were transferred from the original substrate to a silicon substrate with 50\,nm SiO$_2$ layer.

\section{Effect of the growth temperature on the morphology of GaAs and $\mathrm{In}_\mathrm{x}\mathrm{Ga}_\mathrm{1-x}\mathrm{As}$ nanowires}

 We start by discussing the morphology of the nanowires grown under the different conditions. Representative SEM micrographs are shown in figure\,\ref {ingaas:growthtempseries}. GaAs and InGaAs nanowires could be obtained down to a temperature of 560\,$^\circ\mathrm{C}$. By lowering the growth temperature, we observe an increase of the material deposited between the nanowires. These are randomly oriented GaAs deposits that nucleate on the oxide due to the decrease in the mobility of Ga on the surface when the substrate temperature is reduced. Besides the GaAs sample grown at 560\,$^\circ\mathrm{C}$, both the amount of surface deposits and length of the nanowires does not show significant deviations compared to the pure GaAs samples grown under otherwise identical conditions (Figure\,\ref{ingaas:growthtempseries}a-c). This indicates that the sticking coefficient is not significantly altered by the presence of In:Ga with a beam flux ratio of 1:4 as compared to the situation with a pure gallium supply. At 560\,$^\circ\mathrm{C}$, almost the entire surface between the nanowires is covered with randomly oriented (In)GaAs deposits (see figure\,\ref{ingaas:growthtempseries}c,f). The typical size of these structures is up to 500\,nm. Concurrently, the length of the nanowires is reduced with regards to figure\,\ref{ingaas:growthtempseries}a-c. This is also consistent with a reduced surface diffusion of gallium that is incorporating in the surface deposits in the case of low growth temperature. Generally, the observed temperature dependence is in agreement with previous studies for the sticking coefficient of gallium on such SiO$_2$ surfaces \cite{Heiss2008}, which demonstrated a sticking coefficient of unity for temperatures below 565\,$^\circ\mathrm{C}$.

One should note that some of the samples did not show an alignment of the nanowires toward the [1$\overline{\mathrm{1}}$1]B or [11$\overline{\mathrm{1}}$]B  directions of the GaAs (100) growth substrate. We believe this is due to a higher thickness of the oxide, as it was shown elsewhere \cite{Fontcuberta2008}.

\section{Optical properties of the $\mathrm{In}_\mathrm{x}\mathrm{Ga}_\mathrm{1-x}\mathrm{As}$ nanowires}
\begin{figure}[tb!]
	\centering
		\includegraphics{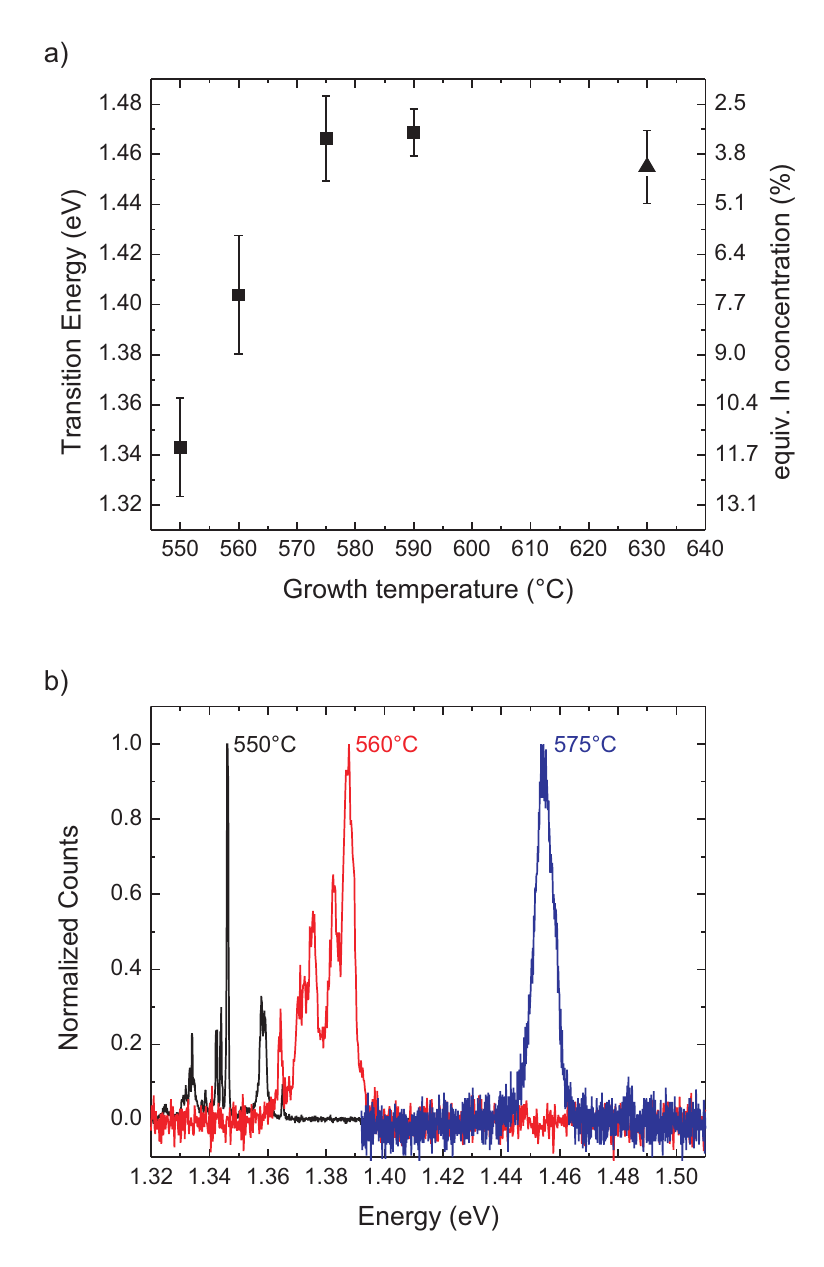}
	
	\caption{\label{ingaas:growthtemp} a) Typical 4.2\,K photoluminescence transitions for $\mathrm{In}_\mathrm{x}\mathrm{Ga}_\mathrm{1-x}\mathrm{As}$ nanowire samples grown at temperatures of 550-590\,$^\circ\mathrm{C}$. \fullsquare~correspond to data from this work while \fulltriangle~corresponds to data from a previous study \cite{Heiss2009} using a higher In deposition rate of 0.088\,{\AA\,s$^{-1}$}.   The right axis of the graph shows the indium concentration of an idealized $\mathrm{In}_\mathrm{x}\mathrm{Ga}_\mathrm{1-x}\mathrm{As}$ material with a band gap related to the transition energy. b) Typical 4.2\,K photoluminescence spectra for $\mathrm{In}_\mathrm{x}\mathrm{Ga}_\mathrm{1-x}\mathrm{As}$ nanowire samples grown at temperatures of 550, 560 and 575\,$^\circ\mathrm{C}$ under otherwise identical conditions.}
	
\end{figure}

\begin{figure}[tb!]
	\centering
		\includegraphics{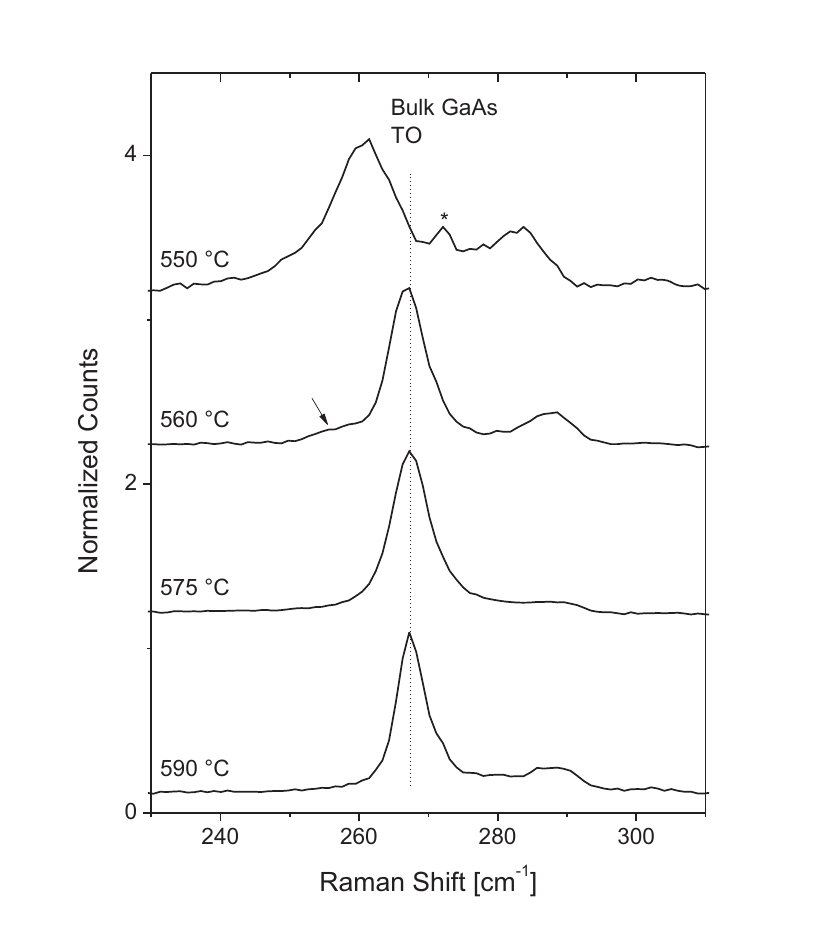}
	
	\caption{\label{ingaas:raman} Raman spectra of single $\mathrm{In}_\mathrm{x}\mathrm{Ga}_\mathrm{1-x}\mathrm{As}$ nanowires from the samples grown at temperatures of 550-590\,$^\circ\mathrm{C}$. The peak marked with (*) corresponds to a line of the laser.}
	
\end{figure}

We expect that the lowering of the growth temperature should have a direct influence on the indium content and therefore on the optical properties. Upon homogeneous incorporation of indium, the band gap E$_g$ should decrease in the form \cite{Goetz1983}:

\begin{equation}
\label{equation1}
\centering
E_g(\mathrm{In}_\mathrm{x}\mathrm{Ga}_\mathrm{1-x}\mathrm{As}) = 1.5192-1.5837x+0.475x^2 (\mathrm{eV})
\end{equation}

Figure\,\ref{ingaas:growthtemp}a summarizes the results of the micro-photoluminescence experiments performed at 4.2\,K on the $\mathrm{In}_\mathrm{x}\mathrm{Ga}_\mathrm{1-x}\mathrm{As}$ nanowires. For completion, we have added the data point of an InGaAs nanowire grown at 630\,$^\circ\mathrm{C}$ from a previous study \cite{Heiss2009}. The graph shows the average photoluminescence peak position obtained from various single nanowires obtained on the same growth run. The error bars indicate the standard deviation of peak positions of the measurements on various nanowires. The indium concentration x in the $\mathrm{In}_\mathrm{x}\mathrm{Ga}_\mathrm{1-x}\mathrm{As}$ nanowires estimated by equation\,\ref{equation1} is indicated in the right axis. We observe a red shift in the  photoluminescence transitions with respect to pure GaAs, which is a clear indication of the presence of indium in the nanowires. Surprisingly, the peak position does not change significantly for nanowires obtained at temperatures from 630\,$^\circ\mathrm{C}$ down to 575\,$^\circ\mathrm{C}$. In all these cases, the optical properties of the nanowires correspond to those of an indium concentration of only $\approx 3\,\%$. As a consequence, we can state that the indium incorporation in the nanowire is not affected by the growth temperature within a variation in this range of temperature. This observation is in quantitative disagreement with thermodynamic calculations of Shen and Chatillon \cite{Shen1990}. This indicates that the incorporation of indium might not be exclusively  governed by a VLS mechanism through the droplet. 
For growth temperatures below 575\,$^\circ\mathrm{C}$ a pronounced redshift of the emission is observed. In principle, this gives an indication that the incorporation of indium in the nanowire may be significantly increased by reducing the growth temperature. However, in order to have a more precise information on the incorporation of indium, we perform a more detailed analysis.

\begin{figure}[tb!]
	\centering
		\includegraphics{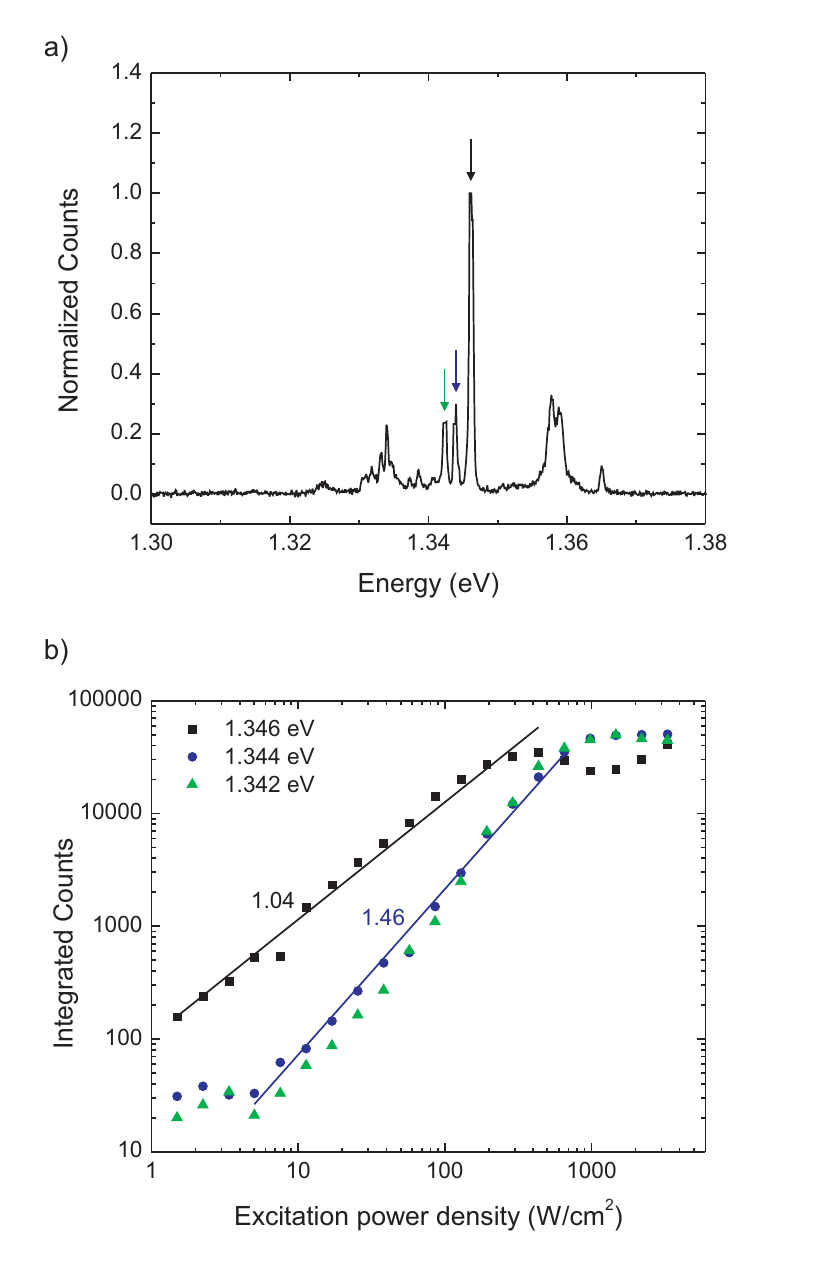}
	
	\caption[4.2\,K photoluminescence spectra of a $\mathrm{In}_\mathrm{x}\mathrm{Ga}_\mathrm{1-x}\mathrm{As}$ nanowire sample grown at 550\,$^\circ\mathrm{C}$]{\label{ingaas:growthtemp550}a) 4.2\,K photoluminescence spectra of a $\mathrm{In}_\mathrm{x}\mathrm{Ga}_\mathrm{1-x}\mathrm{As}$ nanowire sample grown at 550\,$^\circ\mathrm{C}$. b) Excitation power dependence of the three sharp emission lines marked by the colored arrows in a)}
	
\end{figure}

Typical spectra of the InGaAs nanowires obtained under different temperatures are shown in figure\,\ref{ingaas:growthtemp}b. Interestingly, the decrease of the growth temperature is accompanied with an increase of the variations in photoluminescence characteristics from nanowire to nanowire.  At the same time, we also observe that the shape of the photoluminescence spectra changes for growth temperatures below 575\,$^\circ\mathrm{C}$. This is illustrated in figure\,\ref{ingaas:growthtemp}b. Nanowires grown between 630 and 575\,$^\circ\mathrm{C}$ typically exhibit a quite narrow single emission peak around 1.46\,eV. The nanowires grown at lower temperatures present spectra with multiple peaks. Furthermore, the photoluminescence emission of these nanowires is typically inhomogeneous along the nanowire axis. The most extreme example for this was observed in a sample grown at 550\,$^\circ\mathrm{C}$. A photoluminescence spectrum of this sample is shown in figure\,\ref{ingaas:growthtemp550}a. In this case, the emission consists mainly  of three extremely sharp lines. The main emission line at 1.346\,eV has a full width at half maximum of only 365\,$\mu$eV. This spectral width is at the limit of the resolution of our spectrometer. In order to better understand the nature of these transitions, we measured the photoluminescence spectra as a function of the excitation power. The integrated intensity of the lines as a function of the excitation power is shown in figure\,\ref{ingaas:growthtemp550}b. The main emission line at 1.346\,eV has a linear excitation power dependence, while the emission peaks at 1.344\,eV and 1.342\,eV have a nonlinear power dependence as illustrated in figure\,\ref{ingaas:growthtemp550}. Such behavior is typically only observed in low dimensional systems such as quantum dots \cite{Brunner1994}. Interestingly, in order to achieve such a quantum dot like zero dimensional confinement, the related structures must actually be of dimensions much smaller than the diameter of the nanowires. The existence of quantum dots in the nanowire structure may be related to inhomogeneities in the indium concentration at lower temperatures. In order to understand if this is the case, a more detailed analysis of the structure and composition is essential. Therefore, we realized a detailed study on the structure and indium distribution in the nanowires. The results are shown in the following section. 

\section{Structural characteristics of $\mathrm{In}_\mathrm{x}\mathrm{Ga}_\mathrm{1-x}\mathrm{As}$ nanowires}

\begin{figure*}[tb!]
	\centering
		\includegraphics[width=\textwidth]{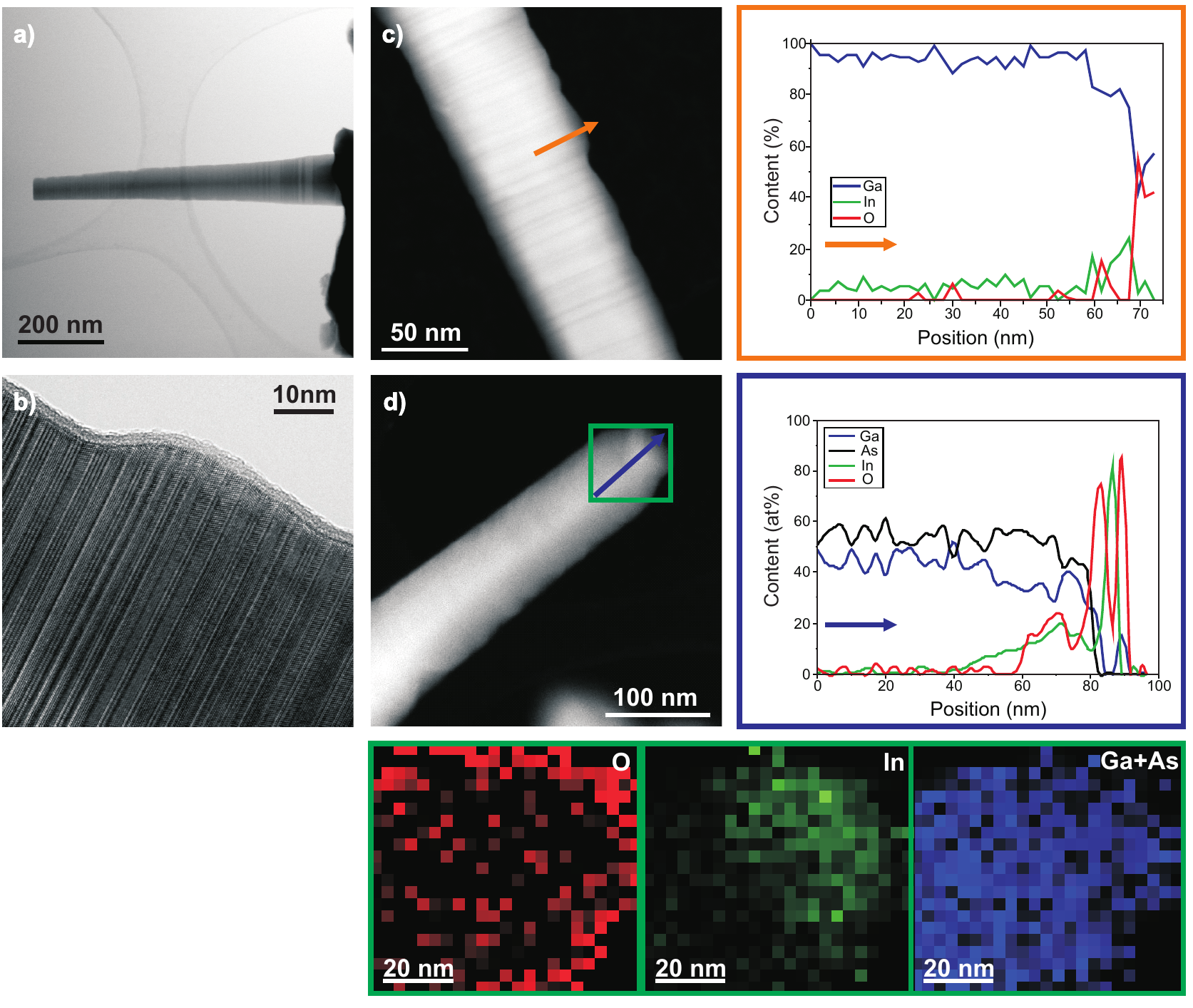}
	
	\caption{\label{ingaas:t550-tem}a) Bright field STEM micrograph of a nanowire from the $\mathrm{In}_\mathrm{x}\mathrm{Ga}_\mathrm{1-x}\mathrm{As}$ sample grown at 550\,$^\circ\mathrm{C}$. b) High resolution micrograph of a nanowire, showing accumulations of material on the surface. c) EELS profile perpendicular to the nanowire growth direction clearly showing higher concentration of indium on the NW side facets. d) EELS profile along the growth direction at the tip of a nanowire. EELS Spectrum Imaging mapping of the tip region, showing a indium rich tip of this nanowire. For the analysis the indium $M_{4,5}$ (443\,eV), the oxygen K-edge (532\,eV), the gallium $L_{2,3}$ edge and the arsenic $L_{2,3}$ edge are used.}

\end{figure*}

\begin{figure}[tb!]
	\centering
		\includegraphics{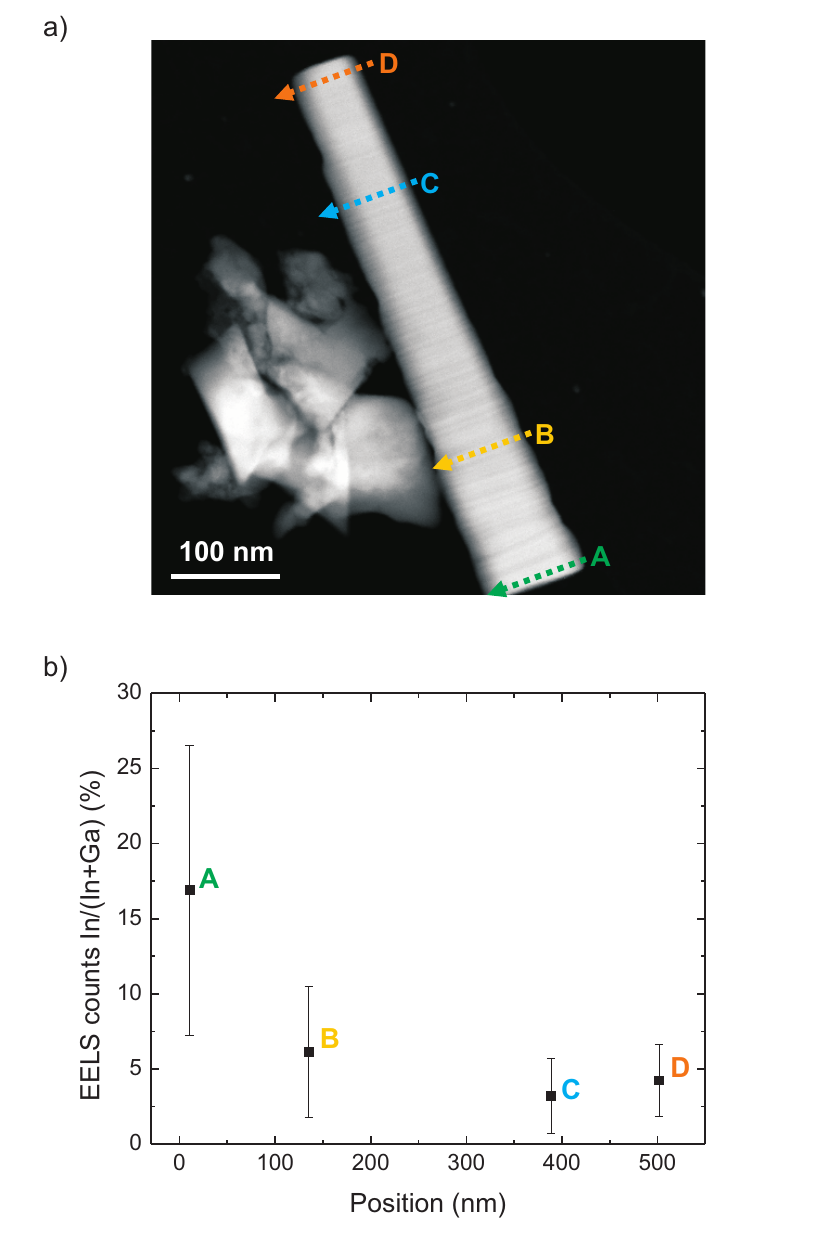}
	
	\caption{\label{ingaas:radialinc} a) ADF-STEM micrograph of a nanowire grown at 550\,$^\circ\mathrm{C}$ b) Indium content obtained from the averaged EELS line scan profiles  A-D marked in a)}

\end{figure}

The frequency positions of the optical phonons in the ternary compound $\mathrm{In}_\mathrm{x}\mathrm{Ga}_\mathrm{1-x}\mathrm{As}$ strongly depends on the compositional fraction x \cite{Brodsky1968}. Therefore, measuring the Raman shift of the optical phonons provides a convenient non-destructive method for estimating the composition. In figure\,\ref{ingaas:raman} Raman spectra of the $\mathrm{In}_\mathrm{x}\mathrm{Ga}_\mathrm{1-x}\mathrm{As}$ nanowires grown at various temperatures are shown. For the sample grown at 590\,$^\circ\mathrm{C}$, the width and the position of the transverse optical mode at 267.5\,$\mathrm{cm}^{-1}$ shows no observable deviations from the pure binary GaAs compound within the experimental uncertainty. By decreasing the growth temperature, we observe a continuous broadening and down shift of the GaAs-like optical phonons. This is directly related to an increase in the indium content in the nanowire. Interestingly, for the nanowire grown at 560\,$^\circ\mathrm{C}$ we observe two overlapping features in the spectra: a broad background denoted by an arrow in figure\,\ref{ingaas:raman} and a superimposed sharper TO mode. It indicates a non-homogeneous distribution of In in nanowire. We suggest that this is due to the formation of a shell with higher In content around the core GaAs nanowire. For the employed laser wavelength of 488\,nm the Raman information depth is in the order of 40\,nm. For the sample grown at 550\,$^\circ\mathrm{C}$, a down shift of the TO mode of 6.9\,$\mathrm{cm}^{-1}$ is observed. From this shift, in comparison with literature data \cite{Islam2002}, we can estimate an indium concentration up to 23\,\% in these nanowires. The spatial resolution of the Raman spectroscopy is not sufficient to obtain information on the axial distribution of indium in these only approximately 700\,nm long nanowires. 

In order to provide a more detailed analysis on spatial inhomogeneities of indium incorporation and at the same time the exact origin of the quantum dot like photoluminescence features observed for the nanowires, the crystalline structure and indium composition of the sample grown at 550\,$^\circ\mathrm{C}$ was analyzed with transmission electron microscopy and EELS. The results are presented in figure\,\ref{ingaas:t550-tem}. Figure\,\ref{ingaas:t550-tem}a shows the bright field STEM micrograph of a typical nanowire of the sample. The nanowires have a length in the order of 700\,nm and exhibit tapering. In figure\,\ref{ingaas:t550-tem}b a high resolution STEM micrograph is shown. The nanowires consist of a highly twinned zinc-blende structure with the sporadic inclusion of some wurtzite segments with thicknesses up to 10\,nm. The surface of the nanowire is oxidized, the oxide thickness being of the order of 2\,nm. Moreover, the side facets of the nanowire are not perfectly flat as we have observed in the past for pure GaAs nanowires \cite{Fontcuberta2008,Uccelli2010}. They present some rounded structures (dome-like), which have a typical height and length respectively of 5\,nm and  50\,nm. In order to obtain more information on the nature of these domes, we realize EELS scans along the diameter of the nanowire at several points. The result is shown in figure\,\ref{ingaas:t550-tem}c. The core of the nanowire is composed of InGaAs with an indium concentration of 3-5\,\%. Interestingly, the indium concentration increases up to approximately 20\,\% at the surface, coinciding with the formation of the nano-domes. We believe that the nano-domes formed at the surface of the nanowire are most likely the origin of the ultra sharp emission lines in the photoluminescence characteristics. We would like to point out that the dimensions of the nano-domes are similar to those observed in Stranski-Krastanov quantum dots \cite{Wasilewski1999}.

Now we would like to discuss the growth mechanisms of the indium rich nano-domes formed on the surface of the nanowires. Figure\,\ref{ingaas:radialinc}a clearly shows that the nanowires grown at 550\,$^\circ\mathrm{C}$ are tapered. This is a clear indication of radial growth. If the formation of indium rich regions is related to the radial growth, one should find a gradient in the indium concentration at the surface. For this purpose, we have realized EELS scans along the nanowire diameter at different points shown in figure\,\ref{ingaas:radialinc}. We have found that indeed the formation of indium rich regions is inhomogeneous along the nanowire and that the concentration is higher at the bottom of the nanowire. Formation of InGaAs quantum dots on GaAs (110) surfaces has been observed in the past for similar growth temperatures \cite{Blumin2006}. Indeed this could account for the sharp features observed in the sample grown at 550\,$^\circ\mathrm{C}$. 

Finally, we have realized an EELS spectroscopy map of the region close to the tip of a nanowire. This is illustrated in figure\,\ref{ingaas:t550-tem}d. At the nanowire droplet at the tip, a peak indium concentration as high as 80\,\% is observed. This clearly shows that the catalyst droplet is very rich in indium during the growth. In the final 40\,nm below the tip, the indium concentration in the solid $\mathrm{In}_\mathrm{x}\mathrm{Ga}_\mathrm{1-x}\mathrm{As}$ gradually increases to a concentration of up to $x \approx 40\,\%$. We believe that the increased incorporation is related to the final part of the growth, when substrate heating is stopped. At that stage, there is still some residual axial growth as the arsenic pressure is maintained. This indicates that an higher indium incorporation during the axial growth might be obtained by even further lowering the growth temperature.

\section{Conclusions}

The formation of $\mathrm{In}_\mathrm{x}\mathrm{Ga}_\mathrm{1-x}\mathrm{As}$ heterostructures in nanowires by catalyst-free molecular beam epitaxy has been studied in dependence of growth temperature. The incorporation of indium in the nanowire core was shown to be limited to 3-5\,\%. A growth temperature series showed that for temperatures below 575\,$^\circ\mathrm{C}$ indium incorporation occurs predominantly through radial growth, as demonstrated by a detailed EELS analysis. The optical properties of such structures resulted in extremely sharp peaks and an excitation power dependence typical of quantum dots.

\section*{Acknowledgments}
The authors acknowledge the experimental support of G. Abstreiter and M. Bichler. Funding through the Marie Curie Excellence Grant SENFED, SFB631, Nanosystems Initiative Munich and ERC Starting Grant Upcon is also greatly acknowledged. This work was partially supported by the Spanish Government projects Consolider Ingenio 2010 CSD2009 00013 IMAGINE and CSD2009 00050 MULTICAT. JA acknowledges the funding from the MICINN project MAT2010-15138 (COPEON) and NanoAraCat. The authors would like to thank the TEM facilities in Instituto de Nanociencia de Arag\'{o}n (INA).

\section*{References}
\providecommand{\newblock}{}

\end{document}